Hot dense capsule implosion cores produced by z-pinch dynamic hohlraum radiation

J.E. Bailey[1], G.A. Chandler[1], S.A. Slutz[1], I. Golovkin[2], P.W. Lake[1], J.J. MacFarlane[2], R.C. Mancini[3], T.J. Buris-Mog[3], G. Cooper[4], R.J. Leeper[1], T.A. Mehlhorn[1], T.C. Moore[5], T.J. Nash[1], D.S. Nielsen[5], C.L. Ruiz[1], D.G. Schroen[6], and W.A. Varnum[1]

1) Sandia National Laboratories, Albuquerque, N.M., 87185-1196

2) Prism Computational Sciences, Madison, WI, 53703

3) Dept. of Physics, University of Nevada, Reno, NV, 89557

4) University of New Mexico, Albuquerque, NM, 87131

5) K-tech Corporation, Albuquerque, NM, 87185

6) Schafer Corp., Livermore, CA, 94550

Hot dense capsule implosions driven by z-pinch x-rays have been measured for the first time. A ~220 eV dynamic hohlraum imploded 1.7-2.1 mm diameter gas-filled CH capsules which absorbed up to ~ 20 kJ of x-rays. Argon tracer atom spectra were used to measure the $T_e$ ~ 1 keV electron temperature and the $n_e$ ~ 1-4 x $10^{23}$ cm$^{-3}$ electron density. Spectra from multiple directions provide core symmetry estimates. Computer simulations agree well with the peak compression values of $T_e$, $n_e$, and symmetry, indicating reasonable understanding of the hohlraum and implosion physics.

Implosion of spherical capsules by x-rays contained in an enclosure known as a hohlraum is the baseline approach for achieving inertial confinement fusion (ICF) [1]. These implosions provide a platform for high energy density physics investigations and may enable studies of matter at conditions similar to stellar interiors. One method for generating the required high energy density x-rays is the z-pinch dynamic hohlraum (ZPDH) [2-4]. Accelerating an annular high-atomic-number z-pinch plasma onto a low-atomic-number low-density cylindrical foam creates this type of hohlraum. The z-pinch plasma impact onto the foam launches a radiating shock that propagates toward the cylinder axis and heats the hohlraum. The z-pinch plasma traps the radiation, improving the symmetry for driving capsule implosions at the center of the hohlraum. In this Letter we describe the first x-ray measurements of hot dense implosion cores produced using z-pinch radiation drive.

ICF requires highly-symmetric capsule implosions. Absorption of 150 kJ of x-ray energy in a carefully-shaped pulse with ~ 1-3 % spatial symmetry is estimated to be sufficient for ignition [1]. The appeal of the ZPDH approach is efficient generation and delivery of x-rays to the capsule. Indeed, the capsules in the experiments reported here absorbed up to 20 kJ of x-rays, about 1/8 of the estimated ignition requirement. However, it remains to be seen whether adequate symmetry and temporal pulse shaping control can be obtained. The experiments reported here represent an important step toward ZPDH symmetry control.

Recent work [5-6] demonstrated reasonable ZPDH performance understanding. The effect of random z-pinch non-uniformities on the shock radiation source was found very small. The systematic asymmetries must now be addressed. The cylindrical ZPDH tends to drive the spherical capsule with equator-hot radiation that, if uncorrected, will produce an implosion elongated along the cylinder axis. Symmetry corrections, including radiation shields, non-cylindrical foams, and local foam dopants, require diagnostics capable of measuring the alteration's effect. Published laser driven implosions [7] showed that x-ray core imaging measurements of the equatorial plane radius (a) compared to the polar radius (b) reflected the x-ray drive asymmetry. Applying this method to the ZPDH is complicated by the difficulty of viewing the implosion through the luminous high-opacity z-pinch plasma (Fig. 1). We consequently employed time- and space-resolved x-ray spectra from tracer atoms in the core to ensure that the measurements reflect the core conditions. This requires two separate spectrometers, one viewing along the polar axis and one viewing in the equatorial plane, since each instrument records only 1D spatial information. The equatorial-to-polar radius ratio can be measured by comparing the two spectra. These spectra lay the foundation for a tomographic reconstruction of the 2-D electron density ($n_e$) and temperature ($T_e$) spatial profiles within the implosion core. In addition, $n_e$ and $T_e$ measurements provide a benchmark of how close the conditions are to ICF requirements and can test whether ZPDH simulations are accurate, since they depend on the drive radiation time history.

In this Letter we describe sample results from two experiment groups. The goal for the first group was single-minded: maximizing the likelihood of measuring core x-ray emission. The experiment representing the first group, Z860, had a 1.7 mm outer diameter, a 36 μm CH wall, and a 4 μm Polyvinylalcohol (PVA) overcoat to retain the fill gas. The fill was 2.8 atm $CD_4$ + 0.085 atm Ar. $CD_4$ was chosen to avoid the difficulty of retaining $D_2$ inside large-diameter thin-CH-wall capsules.

The second group of experiments used a larger capsule and a lower-Z fill gas to increase the Ar emission intensity. This allowed improvements to the polar diagnostics and we implemented a spectrometer viewing along the equator, through the z-pinch plasma. The experiment representing this second group, Z962, used a capsule with a 2.1 mm diameter, a 39 μm CH wall, and a 4 μm PVA overcoat followed by an 800 Angstrom Al overcoat. The fill gas was 13.5 atm $D_2$ + 0.085 atm Ar.

The experiments used the Z facility [2,3] to drive a ZPDH configuration [3,5,8] composed of concentric annular tungsten wire arrays with a 240-wire 40-mm-diameter outer array and a 120-wire 20-mm-diameter inner array. The wire diameter was 7.5 μm. The 14 mg/cc $CH_2$ foam was 6 mm diameter, 10 mm tall on Z860, and 12 mm tall on Z962. Polar line of sight measurements were performed through a 4-mm-diameter aperture that was covered with a 4 μm Ti + 2 μm CH foil to reduce symmetry perturbations. The time-dependent radiation temperature, shock, and capsule radius were diagnosed using x-ray imaging and power measurements [5,8,9]. The ~ 18 Mbar shock had a ~ 35 cm/μsec velocity and the peak radiation drive temperature was ~ 220 eV, prior to the arrival of the shock on axis. The radiation temperature rise time was ~7 ns. The capsule core x-ray emission peaked 2-3 ns before shock arrival on axis and 4-5 ns prior to z-pinch stagnation. The times given below are with respect to z-pinch stagnation, as defined in [8]. The diagnostic suite [10] included capsule implosion time measurements using filtered photoconducting diodes and both activation and time-of-flight neutron detectors. The neutron yield was ~$2 \times 10^{10}$ on Z962, similar to the 1D simulation yield. The Z860 yield was below the ~ $3 \times 10^9$ detector threshold. Detailed comparisons of measured and simulated neutron yields and the dynamic hohlraum radiation temperature and hydrodynamics will be described elsewhere.

Here we emphasize the results obtained from two time- and space-resolved elliptical crystal spectrometers [11] (TREX1 and TREX2) and a time-resolved x-ray pinhole camera (TRPHC1 [10]). The TREX1 spectrometer viewed the capsule along the polar axis (Fig. 1). It used an array of six space-resolving slits to project spectrally-resolved images of the capsule onto six 1-nsec time-gated microchannel plate (MCP) striplines. On Z860 and Z962 the spatial resolution was ~ 330 μm and ~120 μm, respectively. The PET crystal located 2.92 m from the source provided a 1.3 Angstrom spectral range and a spectral resolution of $\lambda/\delta\lambda$ ~ 860. The TREX2 spectrometer viewed the capsule from the side. The instrument components, spectral range, and resolution were the same as for TREX1, except that the source to crystal distance was 4.10 m and the spatial resolution was ~ 200 μm. The data are recorded on Kodak TMAX p3200 film. We unfold the film response using a step wedge, apply the wavelength scale, and account for the relative instrument sensitivity [11]. The TRPHC1 viewed the capsule from the end of the ZPDH, at a 6 degree angle with respect to the ZPDH axis. It used 11 time frames with a 1 ns gate, a 1 ns interframe time, and a ~160 μm spatial resolution.

On Z860 TREX1 recorded Ar spectra in time frames 1 & 2 (Fig. 2). The emission spatial extent was ~ 300 μm, demonstrating that the Ar spectra arise from a well-localized implosion core. A TRPHC1 image recorded simultaneously with the TREX1 frame 1 spectrum (Fig. 3) confirmed that the implosion occurred when the shock in the foam was still at a ~ 1.8 mm diameter. The implosion was therefore radiation driven.

The TREX1 data (Fig. 2) are qualitatively similar to laser hohlraum implosion data [12]. The electron density ($n_e$) was determined from a "line of best fit" analysis [13] of the Stark- and opacity-broadened He-like and H-like Ar line sequences. This analysis determines the $n_{li}\delta r$ product as well as $n_e$, where $n_{li}$ is the lower level population of ion charge state i and $\delta r$ is the plasma size along the line of sight. The $n_e$ inferred from the Z860 TREX1 frame 1 data was $1.3 \pm 0.3$ x$10^{23}$ cm$^{-3}$, corresponding to a mass density $\rho$ ~ 0.35 g/cm$^3$. The compressed fuel $\rho r$ obtained from the $n_{li}\delta r$ product was 9 mg/cm$^2$, under the assumption that the Ar concentration was preserved. The values obtained from the frame 2 data were $n_e$ ~ $3.2 \pm 0.4$ x$10^{23}$ cm$^{-3}$, $\rho$ ~ 0.86 g/cm$^3$, and $\rho r$ ~ 5 mg/cm$^2$. The $n_{li}\delta r$ values lead to Ar line center optical depths of $\tau$ ~ 50-100 for the Heα and Lyα,

$\tau \sim 3$ for the Heβ and Lyβ, and $\tau \sim 0.3 – 1.0$ for the Heγ and Lyγ lines.

An estimate for the radial compression ($C_r$) can be obtained from the ratio of the compressed core density compared with the initial density and assuming a spherical implosion. The frames 1 and 2 TREX1 data correspond to $C_r \sim 5.4$ and $C_r \sim 7.2$, respectively. The actual shape of the implosion is expected to be prolate, since no symmetry corrections were employed here. This is a likely explanation for the small radial size ($C_r \sim 10$) observed in the TRPHC1 image (Fig. 3).

$T_e$ was determined by comparing measurements of spectral line ratios to calculations performed with the non-LTE collisional-radiative SPECT3D model [14]. The model accounts for plasma opacity and our analysis assumed a uniform steady-state Maxwellian plasma. We used both the Lyγ/Heγ and the (He-like satellites to Lyα)/(Li-like satellites to Heα) line ratios. These ratios were selected for their robustness against complications arising from optical thickness and atomic kinetics uncertainties. The electron temperature was $T_e \sim 800 \pm 70$ eV for both frames 1 and 2. In spite of the uniform core approximation, the complete measured frame 1 spectrum agrees well with a synthetic SPECT3D spectrum calculated with $n_e = 1 \times 10^{23}$ cm$^{-3}$, $T_e = 800$ eV, and $\rho r = 5.3$ mg/cm$^2$. However, the high-n lines are not well reproduced in a similar frame 2 comparison. Also, the frame 2 spectrum is ~5x weaker than in frame 1. A preliminary interpretation is that a small residual hot core emits the frame 2 spectral lines through a cool outer layer that has enough ground state Ar population to attenuate the resonance transitions.

The ZPDH radiation and material properties were calculated [6] with 2-D LASNEX [15] simulations that use the multi-group diffusion radiation transport approximation, treating the electrodes as a boundary with a specified time-dependent albedo, and treating the capsule as a loss in the central zones. They also neglect Rayleigh-Taylor instabilities, an approximation justified by Ref. [5,6,8]. The radiation field at the capsule location was calculated by integrating the radiation transport equation within the ZPDH. This radiation was averaged over the capsule surface and used as input for 1-D simulations of the capsule implosion. The post-processed simulation implosion core $T_e$ and $n_e$ are compared with the experimental results in Fig. 4. The simulation relative timing was established using the ZPDH shock trajectory measured using time-resolved x-ray imaging. The simulation $n_e$ and $T_e$ values agree with the experiment at peak core emission intensity. The 0.5 ns timing difference is comparable to the experiment uncertainty. The simulations do not reproduce the residual hot core that is inferred from the frame 2 data, although the $n_e$ rise is very similar to the experiment.

On Z962 spectra were recorded on TREX1 frames 4, 5, and 6 and on TREX2 frame 6 (Figs. 1, 5). The spectral intensity in TREX1 frames 4 and 5 were weaker than the frame 6 spectra by factors of ~ 250 and 10, respectively. The tungsten plasma opacity prevented observation of the weak frame 4 & 5 signals on TREX2. The spatially-averaged core $T_e$ and $n_e$ values determined from the Fig. 5 data using the methods outlined above are shown in Fig. 4. The compressed fuel density at the time of peak emission (frame 6) was $n_e = 1.3 \pm 0.2 \times 10^{23}$ cm$^{-3}$ ($\rho \sim 0.44$ g/cm$^3$), the areal density was $\rho r \sim 8.7$ mg/cm$^2$, and the 1D fuel convergence was $C_r \sim 5.5$. The simulation peak $T_e$ agrees with the data, while $n_e$ is about 0.6 times the experiment value (Fig. 4). The emission peaks approximately 0.5 ns later in the simulation. Also, the data indicate that a relatively small amount of fuel is heated to high temperatures ~2 ns before the emission peaks, a feature not reproduced in the simulations. The absorbed x-ray energy in the Z962 simulations was ~20 kJ, to our knowledge approximately an order of magnitude higher than in prior [1] laser hohlraum implosions. The drive radiation burns through these relatively thin wall capsules when the capsule has imploded to approximately one-half the initial radius. The absorbed energy is dominated by the capsule internal energy, which is proportional to the drive temperature at the peak compression. Increasing (decreasing) the drive temperature will tend to make the implosion occur earlier (later), making the calculated absorbed energy insensitive to drive uncertainties.

Symmetry estimates can be obtained by comparing space-direction lineouts from the two slit-imaged spectra (Fig. 1). Example lineouts taken through the continuum between the Lyδ and Lyγ transitions are shown in Fig. 6. Results of post-processed 2D implosion simulations are superimposed on the measurements. The 2D capsule simulations were driven by the time- and angle-dependent radiation obtained in separate ZPDH simulations [6]. The simulations have a compressed size at peak emission that is larger than the experiment by a factor of ~1.2 (Fig. 6), consistent with the lower simulation electron density. However, the fwhm ratio a/b measured in the polar and equatorial planes is in excellent agreement with the experiment (Fig. 6).

The results demonstrate that the ZPDH can produce diagnosable hot and dense capsule implosions. The large absorbed energy supports the inherent energy efficiency of this approach. Work is in progress to exploit the understanding and the measurement techniques developed here to improve the implosion symmetry and increase the coupled x-ray energy.

We thank the Z accelerator, diagnostics, materials processing, target fabrication, and wire array teams for invaluable and dedicated technical assistance. G. Dunham, G.A. Rochau, and L.P. Mix provided data processing assistance. A General Atomics Inc. team led by D. Steinman fabricated the capsules. We are also grateful to R.J. Leeper, and J.L. Porter, and M.K. Matzen for support and encouragement. Sandia is a multiprogram laboratory operated by Sandia Corporation, a Lockheed Martin Company, for the U.S. Dept. of Energy under contract No. DE-AC04-94AL85000.

(unpublished).

Figure Captions

1. Schematic experiment diagram. The dynamic hohlraum forms when the tungsten z-pinch plasma collides with the foam. TREX1 and TREX2 are spectrometers with spatial resolution directions along the equatorial and polar axes, respectively. The MCP frames record space-resolved spectra at 1 ns time intervals. Ar spectra recorded at the peak emission time are presented from experiment Z962.

2. Spatially-averaged TREX1 Ar spectra from Z860. Synthetic spectra (black) calculated with the SPECT3D model at the specified $T_e$ and $n_e$ are superimposed on the data (red). The times are specified with respect to the z-pinch stagnation.

3. Time-resolved x-ray pinhole image from experiment Z860. This image recorded photon energies above ~ 1 keV. The time is the same as the TREX1 Frame 1 spectrum presented in Fig. 2. The lineout at right corresponds to a horizontal cut through the image.

4. Time-resolved spatially-averaged $T_e$ and $n_e$ measurements (symbols with error bars) compared with post-processed simulations (solid curves). Red, green and blue represent $T_e$, $n_e$, and Ar He$\gamma$ intensity, respectively.

5. Spatially-averaged TREX1 Ar spectra from Z962. Synthetic spectra (black) calculated with the SPECT3D model at the specified $T_e$ and $n_e$ are superimposed on the data (red).

6. Space-direction lineouts through the Z962 data (solid lines with symbols. Post-processed 2-D simulation results (dashed lines) are superimposed on the data. The size of the simulation results was scaled down by a factor of 1.2 to match the equatorial plane fwhm measurement.

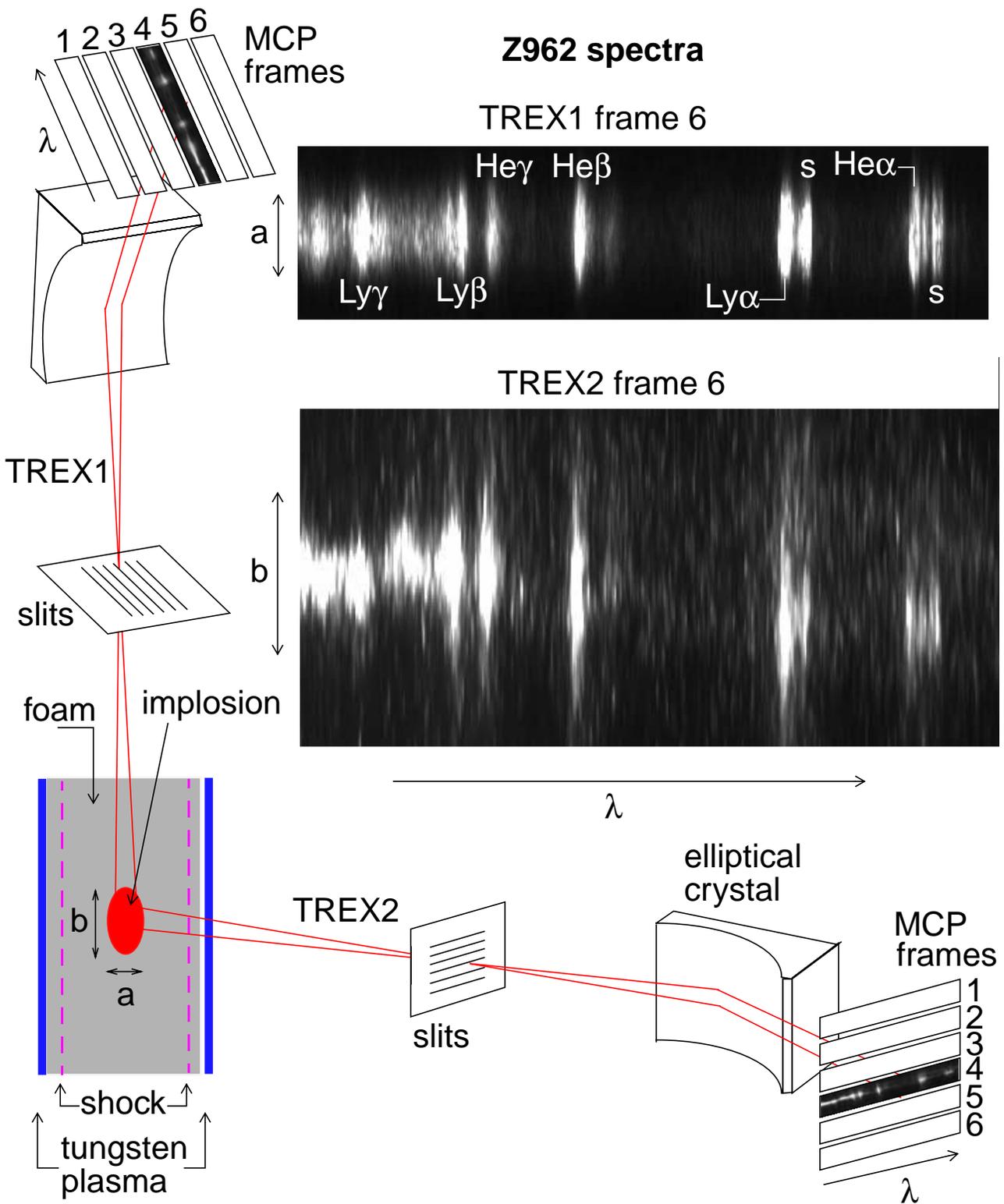

Fig. 1

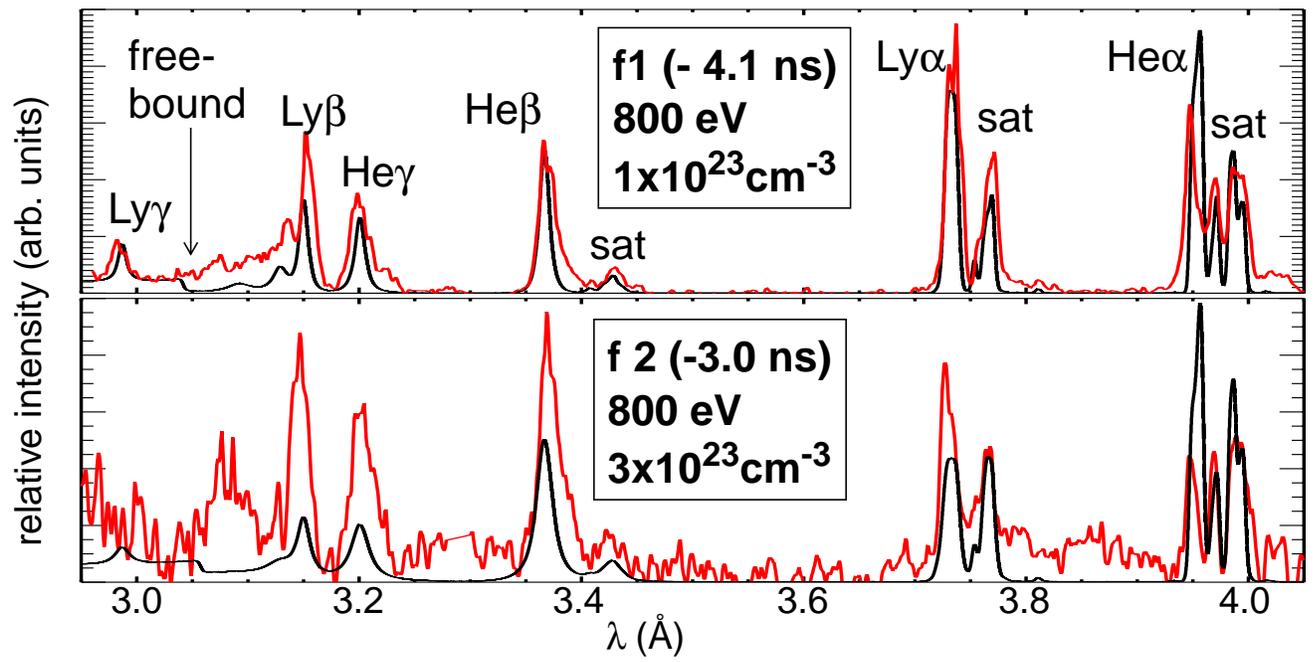

Fig. 2

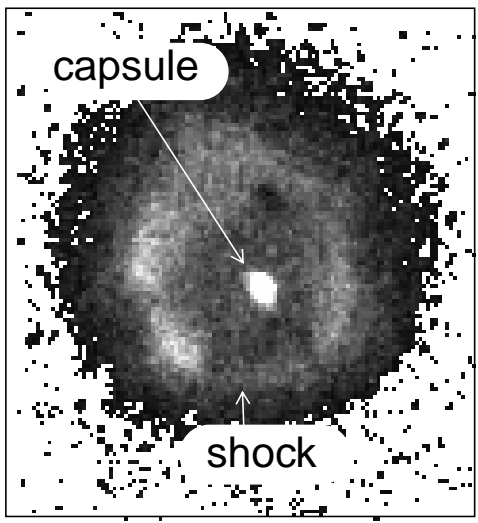 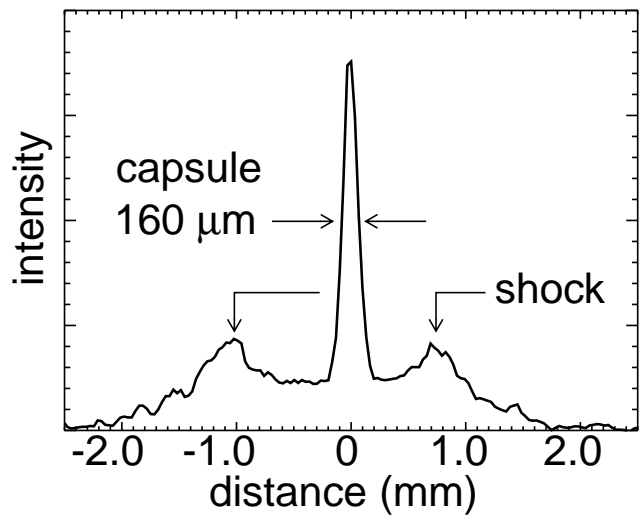

Fig. 3

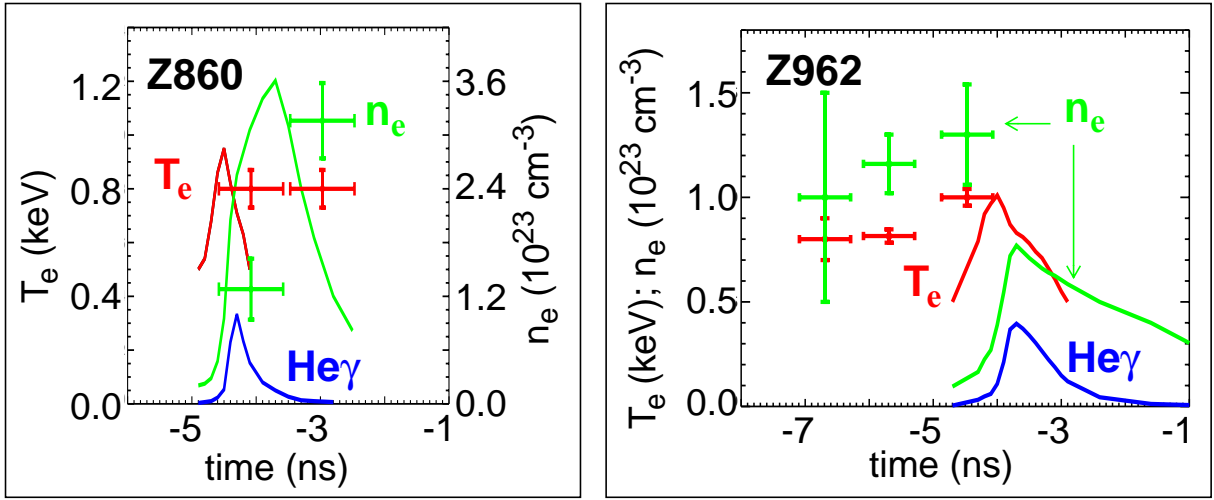

Fig. 4

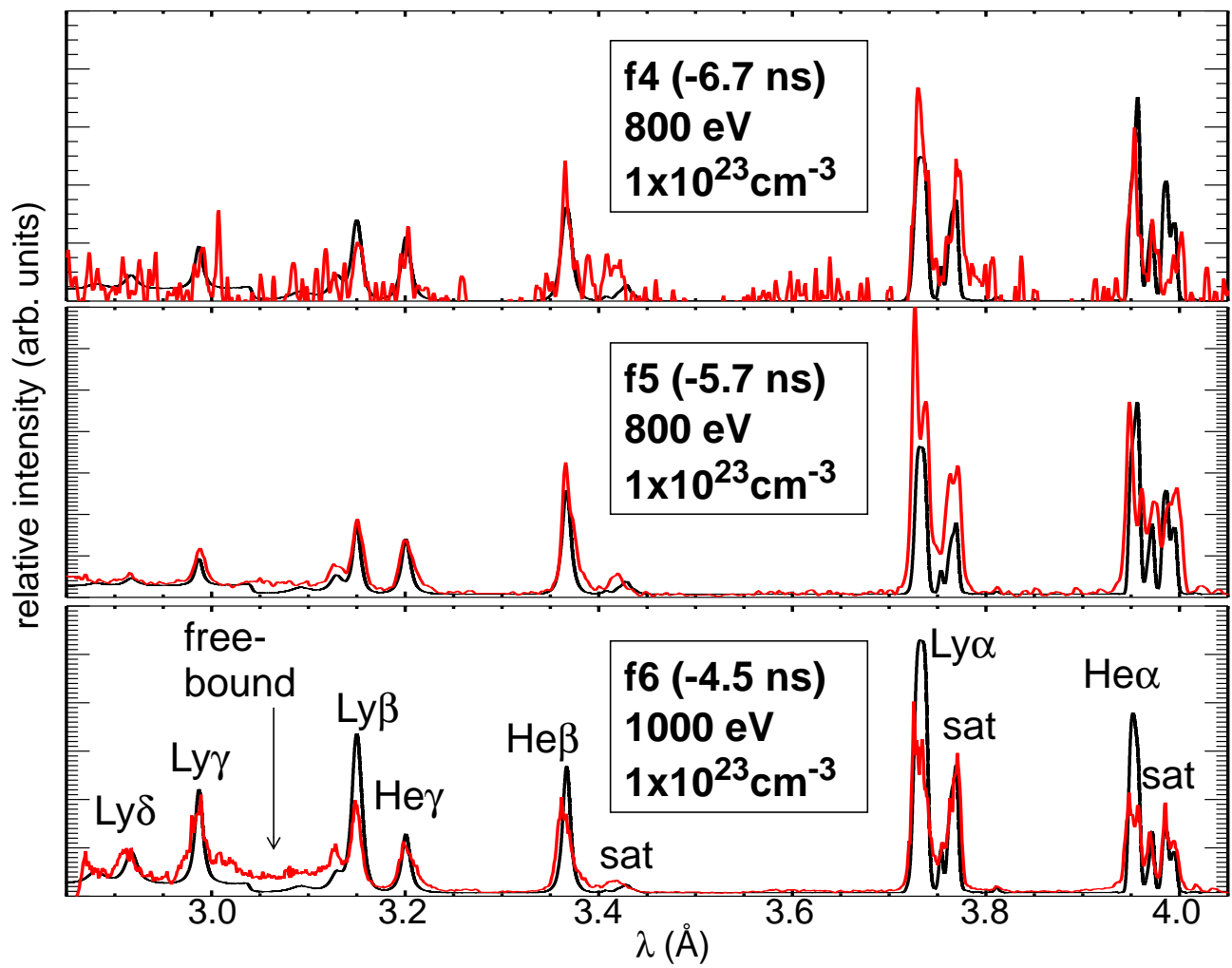

Fig. 5

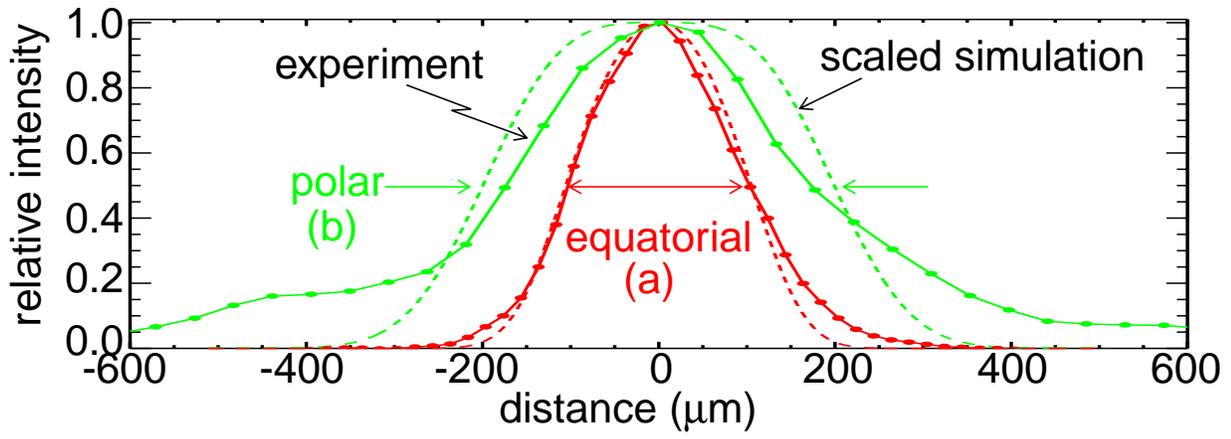

Fig. 6